# Photon-induced switching and tunneling phenomena in a YBCO thin film junction

Xi Yang [1] and A. W. Beckwith [2]

[1]*Fermi National Accelerator Laboratory, P.O. Box 500, Batavia, Illinois 60510-0500*

[2]*Department of Physics, University of Houston, Houston, Texas 77204-5005, USA.*

*77204-5001, USA.*

In the persistent photoconductivity (PPC) phenomenon, illumination of a $YBa_2Cu_3O_{6.5}$ thin film junction with a 1mW He-Ne laser leads to the decrease of the critical voltage (similar to the threshold voltage). The decrease of the critical voltage was reversed by illumination with incandescent light. The critical voltage across the junction was experimentally decreased and increased by alternating illumination between He-Ne laser and incandescent light. We also observed visible quenching of the photo-induced state using a 5mW He-Ne laser. Finally, the threshold behavior of the junction was destroyed by illuminating it with incandescent light.

# I. INTRODUCTION

Photon-induced conductivity changes including Persistent Photoconductivity (PPC) [1] and Photo-induced Superconductivity (PISC) [2] were first reported in oxygen deficient YBCO. PPC and PISC of $YBa_2Cu_3O_{6+x}$ thin films have received special attention [3-9], and the primary reason for this is because subsequent to visible illumination, the photo-induced state of $YBa_2Cu_3O_{6+x}$ films has an increase in conductivity in the normal state (PPC) and an increase in the superconducting transition temperature $T_c$ and diamagnetism in the superconducting state (PISC). It has been suggested that the PISC phenomena has several practical applications such as optically controlled superconductive switches [10].

Two predominant mechanisms have been suggested to explain PPC and PISC in oxygen deficient YBCO: a defect mechanism [11-13] and a photo-induced oxygen ordering mechanism [14]. In the defect mechanism, illumination of a $YBa_2Cu_3O_{6+x}$ film generates electron-hole pairs. The electrons are trapped at oxygen vacancy defects [4,15] in the Cu-O chain layer while the mobile holes are transferred to the $CuO_2$ planes. The effect decays as electrons are released from their traps and recombine with photo-generated holes [13]. In the oxygen-ordering model [8,16], absorbed photons create electron-hole pairs which perturb the local charge and electric field distributions. The local perturbations induce dipole moments that cause oxygen atoms to rearrange within the lattice. These photo-induced changes in the crystal structure result in an increase in the free charge concentration. From optical, transport, and structural experiment results for YBCO obtained by Federici, etc. [17], we believe that the mechanism of PPC should include both defects as well as structural rearrangements. Especially, the trapping of photo-generated electrons at defect sites is a precursor to structural rearrangement.

Unfortunately, there weren't many published photo-experiments concerning the YBCO thin film junction, except the photon-experiment for a superconducting single electron tunneling (SET) transistor.[18]

## II. SAMPLE DESCRIPTION AND EXPERIMENTAL SETUP

$YBa_2Cu_3O_{6.5}$ thin film with a 500 Å thickness was deposited on $SrTiO_3$ substrate by pulsed laser deposition (PLD), and a junction with a 9 μm width was fabricated by ion beam etching. The quality of the junction was carefully examined using scanning electron microscopy (SEM), and no overlapped junction shores were observed. See Fig. 1(a) for the image of the junction by SEM.

[insert figure 1a and then figure 1b about here ]

The standard four-point I-V measurement was used to measure the photo-induced conductivity change of the junction sample, as shown in Fig. 1(b).

## III. EXPERIMENTAL RESULTS AND EXPLANATIONS

The I-V measurement was set in such a way that the minimum current ($I_{min}$) was $-10^{-7}$ A, the maximum current ($I_{max}$) was $10^{-7}$ A, and the output of the current source went from $I_{min}$ to $I_{max}$ in 100 steps, then went backward from $I_{max}$ to $I_{min}$ also in 100 steps. This entire process was called one complete cycle and the frequency of the I-V measurement was 0.005 Hz. Each cycle corresponded to one curve in a diagram. The photo-induced state was initiated by a more than one-hour illumination of a 1mW He-Ne laser at the wavelength of 0.6328 μm. The I-V measurement changed from curve 1 to curve 2 in Fig. 2(a) after a one-hour illumination of the laser light with a photon flux $1.27\times10^{18}$ photons/(s·cm$^2$) and cumulative photon dose $4.6\times10^{21}$ photons/cm$^2$. The laser beam was focused by a 48-mm cylinder lens onto a line slightly out of focus with a 50-μm width at the junction, and the beam was aligned in the direction parallel to the junction. Afterwards, when the cylinder lens was moved toward the sample and the beam was focused onto the junction with a 36-μm width for half an hour, with a photon flux $1.77\times10^{18}$ photons/(s·cm$^2$) and a cumulative photon dose $7.8\times10^{21}$ photons/cm$^2$, the I-V measurement changed from curve 2 to curve 3 in Fig. 2(a).

[insert figure 2a and then figure 2b about here ]

The junction was illuminated in focus for another 20 minutes with a cumulative photon dose $9.9\times10^{21}$ photons/cm$^2$, the I-V measurement changed from curve 3 in Fig. 2(a) to curve 1 in Fig. 2(b). After a regular incandescent light with approximately 15 candlepower at a 10-cm distance to the junction was turned on for 5 minutes with an equivalent cumulative photon dose $1.84\times10^{17}$ photons/cm$^2$ at the wavelength of 555.17 nm, the I-V measurement changed from curve 1 to curve 2 in Fig. 2(b). It was obvious that the YBCO thin film junction changed from the initial state - curve 1 in Fig. 2(a) to final state - curve 1 in Fig. 2(b), then went backward to curve 2 in Fig. 2(b) after a five-minute illumination of the incandescent light. The initial state shown by curve 1 in Fig. 2(a) had a relatively higher resistance within a small current range across the junction than the final state shown by curve 1 in Fig. 2(b). In other words, the critical voltage had a much higher value in the initial state than in the final state. Here, the critical voltage ($V_0$) is defined in a similar way to the threshold voltage, and it is the voltage required for opening a conductive channel across the junction. When the voltage across the junction is below $V_0$, the resistance of the junction is much higher than the resistance when the voltage across the junction is above $V_0$, and $V_0$ is shown in the I-V curve as a discontinuity. This entire process can be explained by the combination of Persistent Photoconductivity (PPC) of oxygen deficient $YBa_2Cu_3O_{6.5}$ thin film since the critical voltage was decreased by the increased cumulative photon dose (from curve 1 in Fig. 2(a) to curve 1 in Fig. 2(b)), and Infrared Quenching of the photo-induced state since the critical voltage was increased by the illumination of incandescent light (from curve 1 in Fig. 2(b) to curve 2 in Fig. 2(b)). What really happened to the YBCO junction sample in the above process was that the photo-generated electrons were trapped at defects-presumably oxygen vacancies, and the trapping of electrons prevented the recombination of holes with electrons, resulting in the transfer of holes to the $CuO_2$ planes where they contributed to the conductivity. The photo-induced state was continuously initiated by a more than one-hour illumination using the 1mW He-Ne laser with a cumulative photon dose about $10^{22}$ photons/cm$^2$, which is close to what was observed by V. I. Kudinov, etc.[5] During the illumination, the temperature of the YBCO junction was monitored by a thermocouple, and the maximum change in temperature was less than one and a half degrees (Celsius). The entire experiment was run at room temperature. The resistance across the YBCO junction decreased from curve 1 in Fig. 2(a) to curve 1 in Fig. 2(b). This was mainly caused by the photo-induced state, and this maximum decrease in the resistance across the junction happened only when the trapping of photo-generated electrons reached saturation and photo-generated electrons filled all the oxygen vacancies in the YBCO film. Afterwards, the resistance of the YBCO junction was increased by the illumination of the incandescent light. In this situation, the only possible explanation was IR quenching of the photo-induced state. It is easy for us to understand that if photo-generated electrons are trapped at defects such as oxygen vacancies, it is possible to reverse the effect by photo-exciting the electrons out of the defects, and once the electrons are freed, they could recombine with mobile holes and quench the persistent photoconductivity. Experimentally, Federici, etc. observed that the PPC and PISC state could be reversed by using an infrared light (IR) source to photo-ionize trapped electrons. The photon energy was chosen because it was close to the 1-eV energy barrier inferred from their thermal decay measurement. During the experiment, the freed electrons recombined with holes, thereby partially quenching the photo-induced persistent conductivity state. In our situation, we used the incandescent light to illuminate the YBCO junction sample after the photo-induced state and observed IR quenching of the photo-induced state from curve 1 in Fig. 2(b) to curve 2 in Fig. 2(b). This switching phenomenon, as shown in Fig. 2 (from Figs. 2(a) to 2(b)), was repeated three times later at the same experimental conditions.

After we observed the switching phenomenon, the 1mW He-Ne laser remained in focus for another five hours with a cumulative photon dose $3.2\times10^{22}$ photons/cm$^2$. The I-V measurement changed to Fig. 3.

[insert figure 3 about here ]

The critical voltage was decreased further and the junction had a positive bias when the current across the junction was zero. For this dramatic decrease in the critical voltage, it was most likely that this five-hour illumination created more oxygen vacancies in the YBCO thin film. These oxygen vacancies can trap more photo-generated electrons and allow more mobile holes to be transferred to the $CuO_2$ plane layer. Since there were more photo-generated charge carriers in the YBCO film now than before, the resistance of the YBCO junction was further decreased. The zero-current voltage across the junction was caused by the unequal oxygen vacancies in YBCO films on both sides of the junction. The misalignment of the focused laser beam across the junction caused the laser energy density to be different on the two sides of the junction, so the density of the oxygen vacancies was also different. When we switched on the incandescent light, the I-V curve changed dramatically. The discontinuity of the I-V curve disappeared, and there was only a hysteresis behavior in the I-V measurement. Here, the hysteresis behavior is shown in the difference between the part of the I-V curve when the current was increased from the minimum to the maximum and the part of the I-V curve when the current was decreased from the maximum to the minimum, as shown in Fig. 3. The reason why the discontinuity in the I-V curve disappeared was because the incandescent light freed more electrons from trapping centers (oxygen vacancies) and the free electrons in the $I_{out}$ side (shown in Fig. 1(b)) of the YBCO film directly contributed to the current without crossing the junction and losing the energy for overcoming the junction barrier. In the situation that free electrons in the $I_{out}$ side were not sufficient for the setting current, some of the free electrons in the $I_{in}$ side were required to cross the junction and contribute to the setting current. The movement of a charge crossing the junction costs energy and contributes to the critical voltage. Once photo-generated free charges were sufficient to provide the current, none of the charges contributing to the current needed to cross the junction and overcome the energy barrier. Therefore, the critical voltage in the I-V measurement decreased to zero.

Twenty-four hours later, before illumination, the critical voltage of the I-V measurement increased, as shown in Fig. 4(a) curve 1.

[insert figure 4a and figure 4b about here ]

We used a 2.5 mW He-Ne laser, and the laser beam was focused onto the junction by the same cylinder lens. One and a half hours later, the I-V measurement was as shown by curve 2 in Fig. 4(a) after the applied cumulative photon dose $2.39\times10^{22}$ photons/cm$^2$. The critical voltage value kept on increasing. Afterwards, we increased the power of the He-Ne laser to 5mW by decreasing the attenuation. The YBCO junction was continuously illuminated by the 5mW laser energy for another one and a half hours with a photon flux $8.85\times10^{18}$ photons/(s·cm$^2$) and a cumulative photon dose $7.2\times10^{22}$ photons/cm$^2$, and simultaneously the I-V measurement kept going. Nothing happened until the current went backward from $10^{-7}$ A to $-4.18\times10^{-8}$ A. Here the voltage value suddenly jumped from 0.00 V to –0.19 V, as shown in Fig. 4(b). Afterwards, the I-V measurement remained the same at the same illumination condition. After we switched off the laser beam, the critical voltage increased slowly, as shown in Fig. 4 (c). The only possible explanation for the laser inducing a critical voltage jump observed in Fig. 4(b) was that the 5mW He-Ne laser with a visible wavelength also released the trapped electrons and caused the recombination of electron-hole pairs. This process was similar to IR quenching of the photo-induced state except that the illumination in Fig. 4(b) had a visible wavelength. This also can be used to explain that the critical voltage in Fig. 4(a) was increased by the illumination of the 2.5mW He-Ne laser. Up to now, we observed visible quenching of the photo-induced state except it took a longer time using a visible light than using IR light with a matched wavelength for de-trapping. This might be explained by the excitation of the trapped electrons with a matched IR wavelength having a much higher efficiency than a visible light even if the visible light has a higher energy per photon.

Without illumination for twelve hours, the critical voltage dropped again. This was because the oxygen vacancies trapped electrons and generated mobile holes in the $CuO_2$ plane to recover the photo-induced state. The zero-current voltage was induced by the different concentrations of oxygen vacancies in the YBCO films on both sides of the junction. According to the observation in Fig. 3, if we varied the dose of the illumination from the incandescent light, we should be able to observe that the discontinuity of the junction smeared out gradually. In other words, the critical voltage should be a function of the illuminating dose from the incandescent light. The advantage of using the incandescent light was that the spectrum covers a range from near UV to near IR, so the ability to excite the trapped electrons out of the oxygen vacancies is much higher than with the He-Ne laser. If there were sufficient free electrons generated by the incandescent light, they would provide the current without crossing the junction and there would be no cost of energy needed for overcoming the barrier. Depending upon the free charge concentration, if there were enough free charges in the YBCO film to form the current, none of the charges contributing to the current needed to overcome the energy barrier in order to cross the junction and in this situation the critical voltage was zero and the discontinuity disappeared. If the current was partially contributed by the free electrons in the YBCO film without crossing the junction and partially from the electrons crossing the junction, the critical voltage was proportional to the fraction of the charges crossing the junction. As shown in Fig. 5, when the incandescent light was moved toward the YBCO films from 40 cm to 5 cm such that the illuminating dose was increased, we did observe that the critical voltage decreased gradually during this movement.

[insert figure 5 about here ]

When the distance between the incandescent light and the junction sample was around 5 cm, the discontinuity of the I-V curve almost disappeared and this was equivalent to the zero value of the critical voltage.

## IV. CONCLUSION

We observed a photo-induced switching phenomenon in an oxygen deficient YBCO thin film junction. The critical voltage of the I-V measurement decreased dramatically (50%) due to the illumination of the junction by a 1mW He-Ne laser for more than one hour, and the change of the critical voltage could be explained by Persistent Photoconductivity (PPC) [1]. The photo-induced state was quenched by IR radiation (incandescent light), and this process was shown in the I-V measurement by the recovery of the initial I-V behavior. The entire process including the photo-induced state and IR quenching of the photo-induced state was shown in the I-V measurement by the I-V curve changing from the initial state (with a large critical voltage) to the final state (with a comparably small critical voltage) and finally changing back to the initial state. Also this phenomenon was experimentally repeatable. So we called this experimental observation the photo-induced switching phenomenon.

We also observed the visible light quenching of the photo-induced state by illuminating the junction with a 5mW He-Ne laser at a photon flux $8.85\times10^{18}$ photons/(s·cm$^2$) . This is quite different from what was observed previously.

Finally, we observed that the threshold behavior of the I-V measurement was gradually smeared out by increasing the radiation dose from an incandescent light. The de-trapping of the electrons from oxygen vacancies of the YBCO film attached by the outgoing side of the current source provided sufficient charge carriers to contribute to the current, instead of the free electrons on the other side of the junction tunneling through the junction to contribute to the current. Thus, the energy cost for electrons tunneling through the barrier was spared by the increase of the free electron density, and the result of this was the I-V threshold behavior smeared out. We called it the quasi-tunneling phenomenon. Since when the junction was illuminated either by a He-Ne laser or by a incandescent light, the current in the I-V measurement was the sum of contributions from the free electrons in the YBCO film without crossing the junction and from the electrons crossing the junction, it is difficult for us to separate these two contributions in the above experimental configuration and extract the relationship between the photo-doping and tunneling barrier height. A future experiment should be implemented for the purpose of separating these two contributions such that the information of how the photo-doping influences the tunneling barrier height can be obtained.

The switching and quasi-tunneling phenomena can be potentially used in optically controlled switching devices and integrated voltage generators due to the comparable large photo-induced voltage change across the junction.

## FIGURE CAPTIONS

FIG. 1. (a) YBCO junction image taken by Scan Electron Microscopy (SEM). The vertical strip (indicated by a horizontal arrow) represents the junction that has a 9-μm width, and the YBCO thin film covers both sides of the junction; (b) Schematic diagram of the standard four-point I-V measurement. The setting current goes in the sample from the point represented by $I_{in}$ and comes out from the point represented by $I_{out}$. The voltage meter measures the voltage across $V_+$ and $V_-$.

FIG. 2. The I-V measurement runs from $-10^{-7}$ A to $10^{-7}$ A with 100 data points, and backward from $10^{-7}$ A to $-10^{-7}$ A with 100 data points, each data point takes one second. (a) Curve 1 was measured before illumination. Curve 2 was measured after the junction had been illuminated for one hour by a 1mW He-Ne laser, which was focused by a 48-mm cylindrical lens to a line with a width of 50 μm onto the junction. Following curve 2, curve 3 was measured after the junction had been illuminated for half an hour by a laser beam focused onto the junction with a width of 36 μm; (b) curve 1 was measured after another 20 minutes illumination following curve 3 in FIG. 2(a) and in the same illumination conditions, and curve 2 was measured after a several-minute illumination of incandescent light.

FIG. 3. The next day, the solid curve was measured before illumination. The curve represented by square points was measured when the junction was illuminated by incandescent light. The dashed curve was measured after the incandescent light was switched off.

FIG. 4. The day afterward: (a) the black curve was measured before illumination. The curve represented by open circles was measured after a one-hour illumination of 2.5mW

He-Ne laser with a 36-μm beam width focused on the junction; (b) continuously illuminating the junction with 5mW He-Ne laser after FIG. 4(a) in the same focused condition, the I-V measurement stayed in the same curve represented by solid squares. After one and half an hours' illumination, the voltage across the junction suddenly jumped to a much larger value of about −0.24 volts at a current of $-4.9 \times 10^{-7}$ A, and remained in the curve represented by open triangles even if the laser was kept on; (c) when we switched off the laser, the curve changed from the one represented by solid squares to the one represented by open circles.

FIG. 5. (a) When the current was increased (indicated by the solid curve with arrow), there was no light. Afterwards, an incandescent light was kept a distance of 40 cm from the junction when the current was running backward; (b) the curve represented by open squares was measured when the distance between the junction and the light was 15 cm, and the curve represented by solid circles was measured when the distance between the junction and the light was 5 cm.

(a)

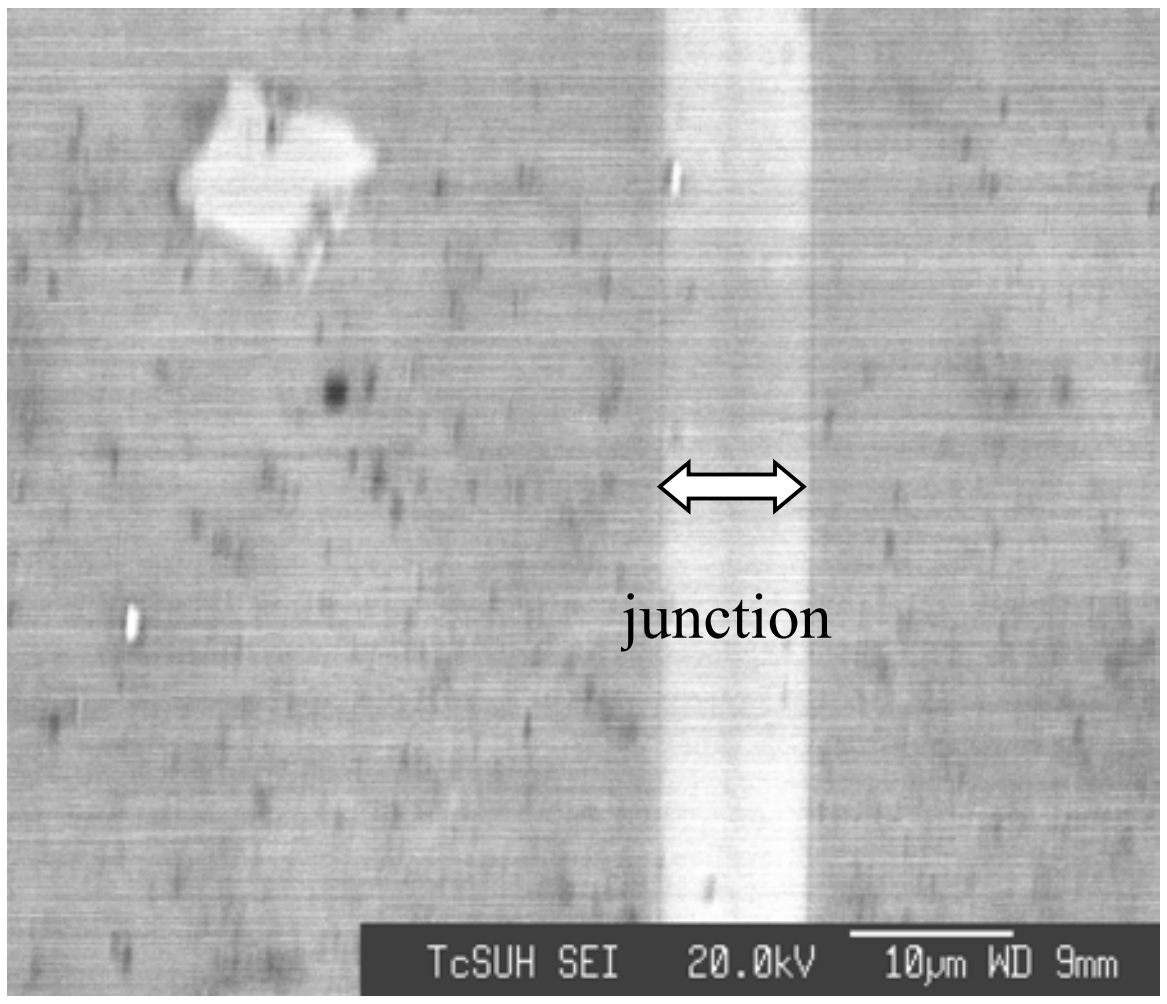

(b)

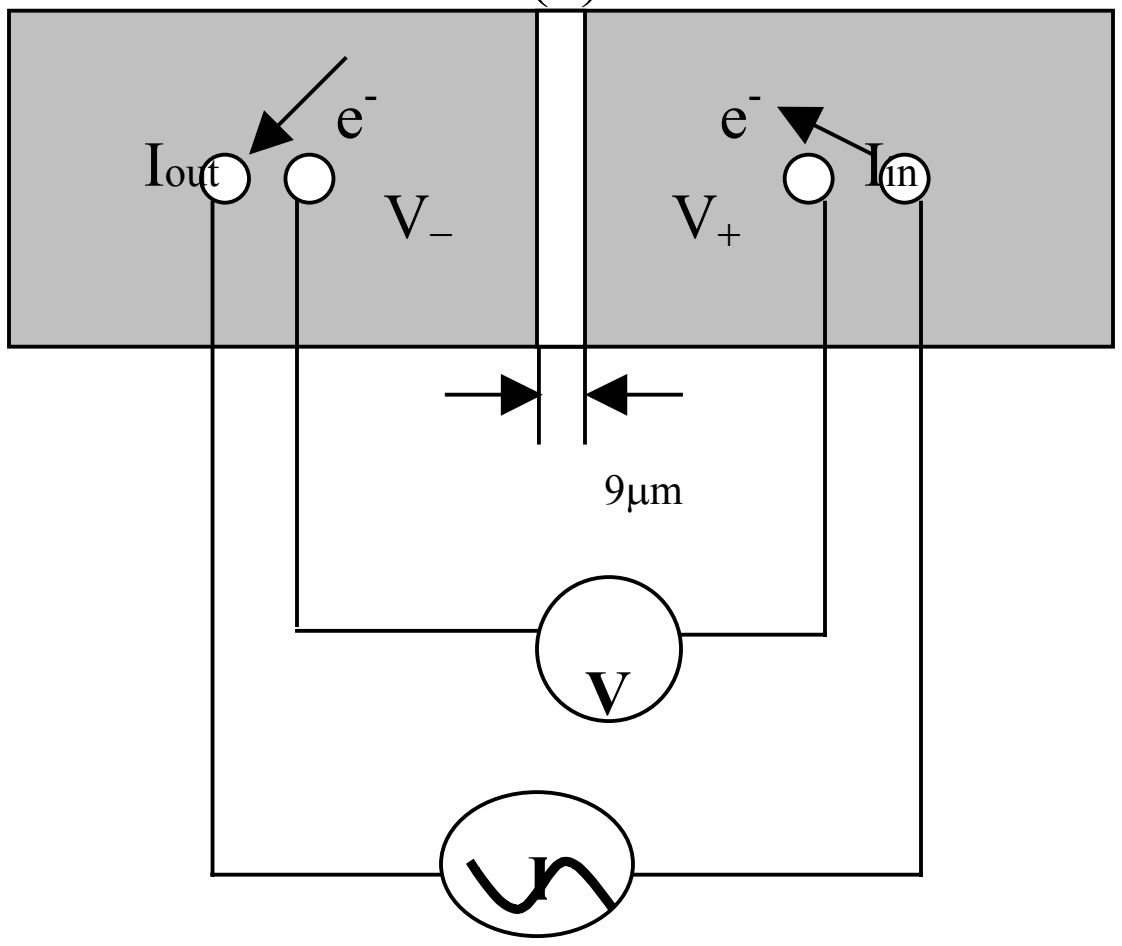

FIG. 1.
Yang et al.

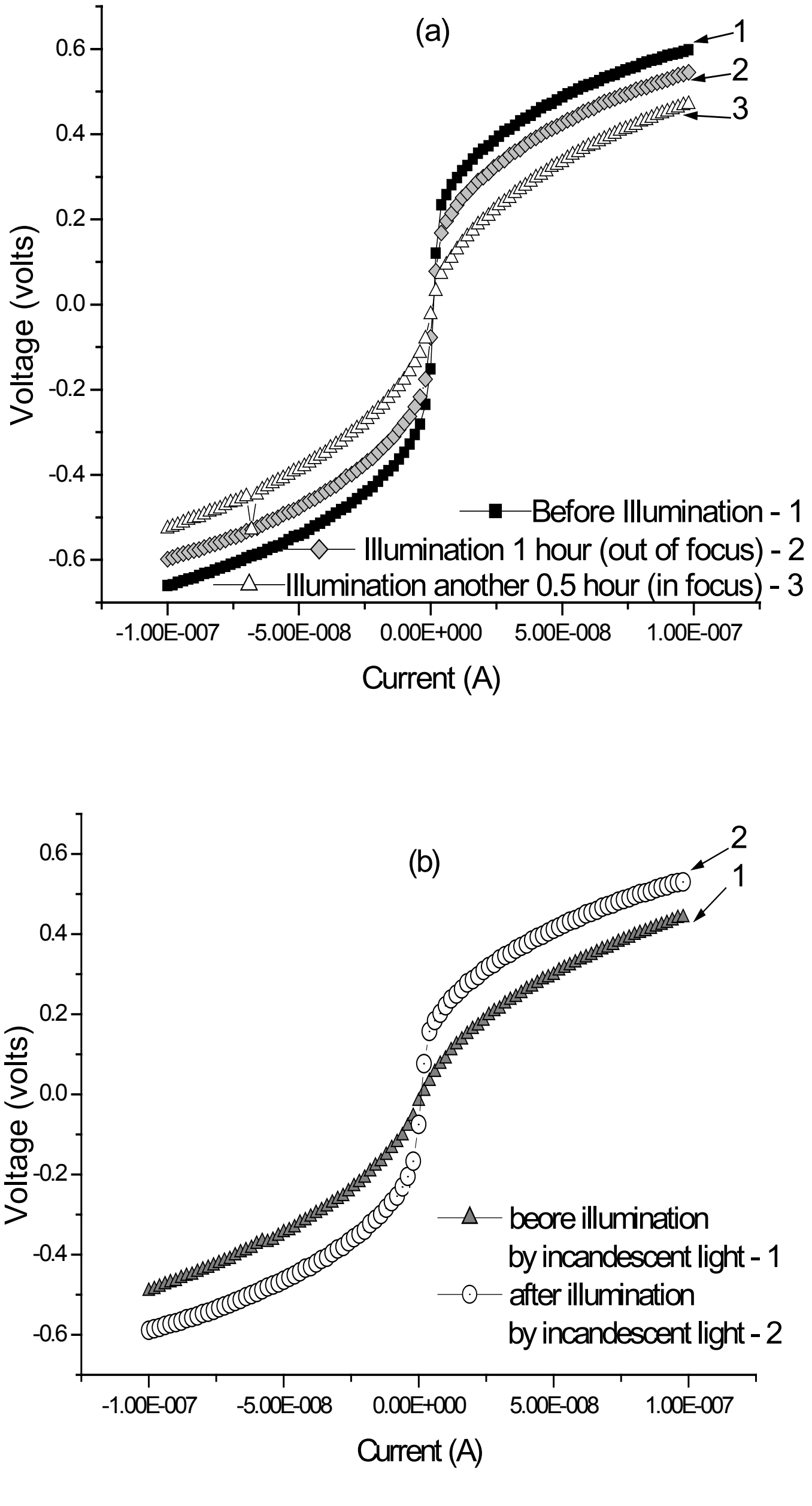

FIG. 2.
Yang et al.

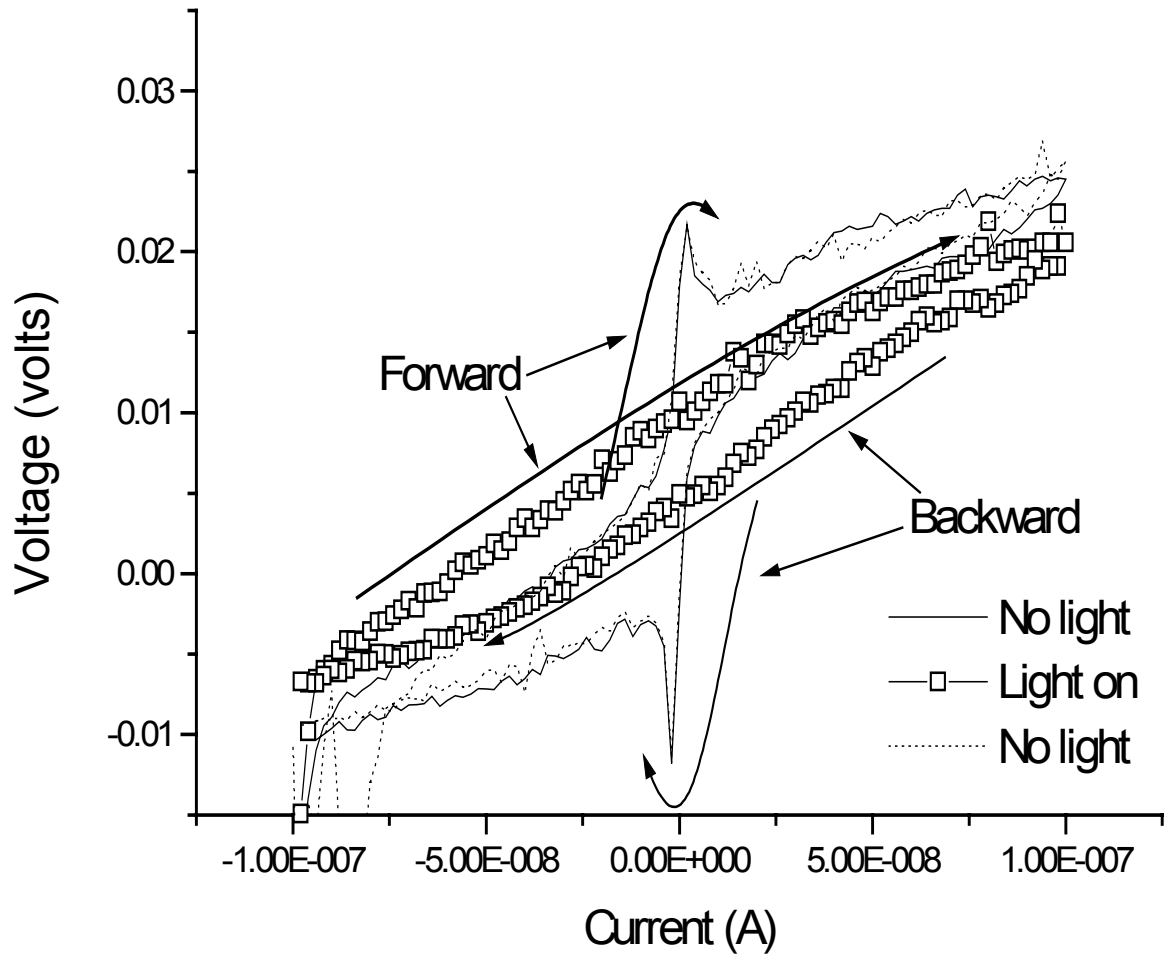

FIG. 3.
Yang et al.

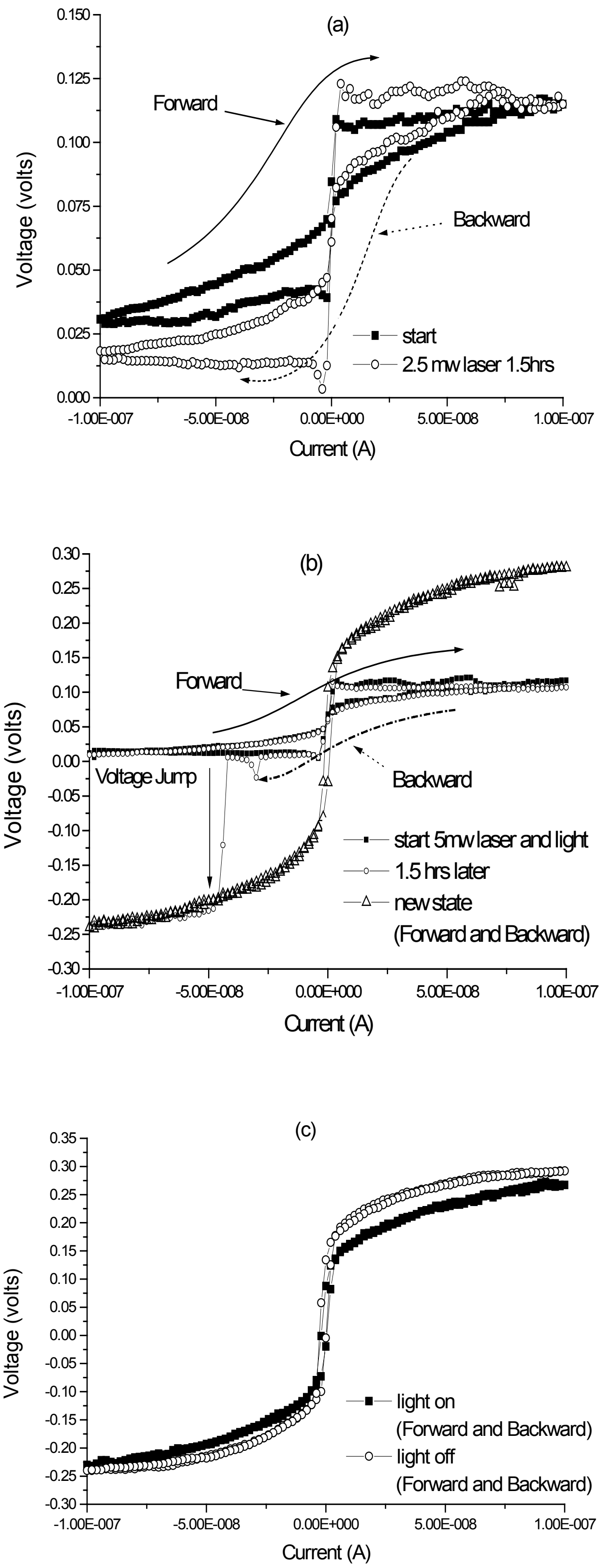

FIG. 4.
Yang et al.

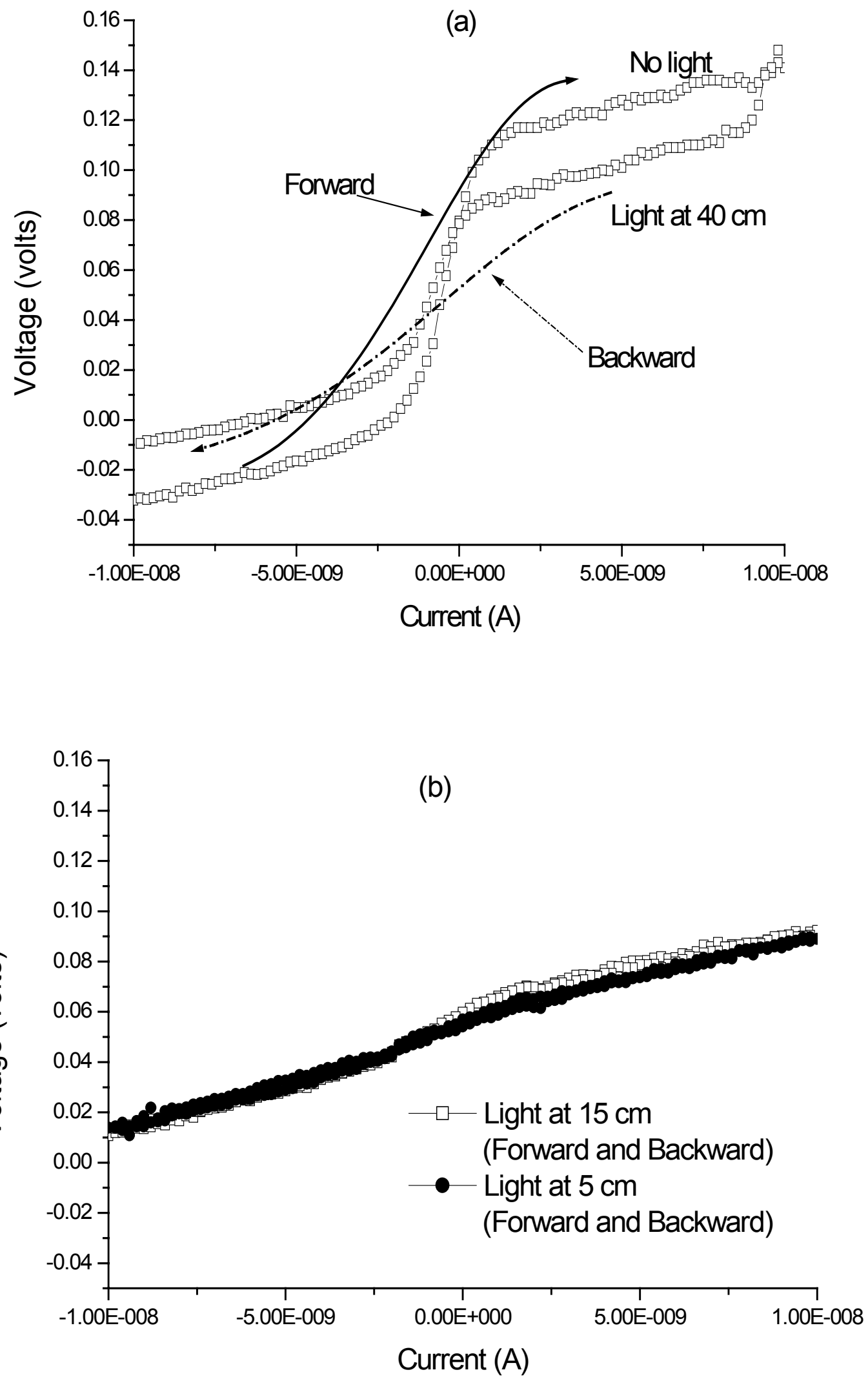

Figure 5

Yang et al.